\documentclass{aa}
\usepackage{graphicx}
\usepackage{natbib}

\usepackage{txfonts}
\begin{document}
\title{Direct detection of exoplanet host star companion $\gamma$ Cep B
and revised masses for both stars and the sub-stellar object}

\author{Ralph Neuh\"auser \inst{1} \and
Markus Mugrauer \inst{1} \and
Misato Fukagawa \inst{2,3} \and
Guillermo Torres \inst{4} \and
Tobias Schmidt \inst{1}
}

\institute{
Astrophysikalisches Institut, Universit\"at Jena, Schillerg\"asschen 2-3, 07745 Jena, Germany \and
Spitzer Science Center, California Institute of Technology, Pasadena, CA 91125, USA \and
Division of Particle and Astrophysical Sciences, Nagoya University, 
Furo-cho, Chikusa-ku, Nagoya 464-8602, Japan \and
Harvard-Smithsonian Center for Astrophysics, 60 Garden Street, Cambridge,  MA 02138, USA
}

\offprints{Ralph Neuh\"auser, \email{rne@astro.uni-jena.de}}

   \date{Received 17 Oct 2006; accepted 27 Oct 2006}

  \abstract
{The star $\gamma$~Cep is known as a single-lined spectroscopic triple system 
at a distance of 13.8 pc, composed of a K1 III-IV primary star with $V = 3.2$ mag, 
a stellar-mass companion in a 66--67 year orbit (Torres 2006), 
and a substellar companion with $M_p \sin i = 1.7$~M$_{\rm Jup}$ 
that is most likely a planet (Hatzes et al. 2003).
}
   {We aim to obtain a first direct detection of the stellar companion,
to determine its current orbital position (for comparison with the
spectroscopic and astrometric data), its infrared magnitude and, hence, mass.
}
  {We use the Adaptive Optics camera CIAO at the Japanese 8m telescope Subaru
on Mauna Kea, Hawaii, with the semi-transparent coronograph to block most
of the light from the bright primary $\gamma$ Cep A, and to detect at the
same time the faint companion B. In addition, we also used the IR camera $\Omega$ Cass
at the Calar Alto 3.5m telescope, Spain, to image $\gamma$ Cep A and B by 
adding up many very short integrations (without AO).
}
   {$\gamma$ Cep B is clearly detected on our CIAO and $\Omega$ Cass images.
We use a photometric standard star to determine the magnitude of B after PSF
subtraction in the Subaru image, and the magnitude difference between A 
and B in the Calar Alto images, and find an average value of $K = 7.3 \pm 0.2$ mag.
The separations and position angles between A and B are measured on 15 July 2006 
and 11 and 12 Sept 2006, B is slightly south of west of A.
}
   {By combining the radial velocity, astrometric, and imaging data, 
we have refined the binary orbit and determined the
dynamical masses of the two stars in the $\gamma$~Cep system, namely
$1.40 \pm 0.12$~M$_{\odot}$ for the primary and $0.409 \pm
0.018$~M$_{\odot}$ for the secondary (consistent with being a M4
dwarf). We also determine the minimum mass of the sub-stellar companion
to be $M_p \sin i = 1.60 \pm 0.13$~M$_{\rm Jup}$.
}

\titlerunning{Direct detection of $\gamma$ Cep B}


   \maketitle

\section{Introduction: $\gamma$ Cep}

The bright star $\gamma$ Cep (also HD 222404 or HIP 116727) located near the
north celestial pole ($\alpha = 23^{\rm h} 39^{\rm m} 20.8^{\rm s}$ and 
$\delta = +77^{\circ} 37' 56.2''$ for J2000.0)
is known as a single-lined spectroscopic triple:
The primary star has spectral type K1 III-IV, is visible to the naked eye ($V=3.2$ mag)
and is located at a distance of 13.8 pc.
A low-mass stellar companion with an orbital period of several decades 
was discovered spectroscopically by Campbell et al. (1988), 
but until now it has never been imaged directly 
(e.g. Hatzes et al. 2003, Mugrauer et al. 2006).
Campbell et al. (1988) and Walker et al. (1992)
discussed the evidence for additional radial velocity variations
indicative of a very low-mass, possibly planetary companion. Hatzes et
al. (2003) confirmed that there is indeed a third object in the
system with a minimum mass of $M_p \sin i = 1.7$~M$_{\rm Jup}$, i.e.,
probably a planet, with a semi-major axis of 2 AU and a 900-day
eccentric orbit ($e = 0.12$).

Several investigators have attempted to constrain the orbit as well
as the mass of $\gamma$~Cep~B (the stellar companion) on the basis of
available spectroscopic data. The estimated period of the binary from
these studies has ranged from about 30 to 66 yr (Walker et al.
1992, Griffin et al. 2002).
Most recently, Torres (2006) put together not only the spectroscopic data
(dating back to 1902), but also {\it Hipparcos} and ground-based astrometric data 
(dating back to 1898) and obtained new results 
for the primary star ($\gamma$ Cep Aa), 
the secondary (B), and the sub-stellar companion (Ab):
B orbits A in an eccentric orbit ($e=0.4085 \pm 0.0065$) with
$66.8 \pm 1.4$ year period and a semi-major axis of $19.02 \pm 0.64$ AU;
the primary has a mass of $1.18 \pm 0.11$~M$_{\odot}$,
temperature of $4800 \pm 100$ K, and an age of 6.6 Gyrs;
the unseen stellar companion should have a mass 
of $0.362 \pm 0.022$~M$_{\odot}$, i.e. spectral type about M4; 
the minimum mass of the lowest-mass component in the system 
is then $M_p \cdot \sin i = 1.43 \pm 0.13$~M$_{\rm Jup}$;
the astrometric data yield an upper mass limit of 17~M$_{\rm Jup}$
(at the 3 $\sigma$ confidence level), i.e. certainly sub-stellar, 
probably planetary; all values above are from Torres (2006).
The predictions for $\gamma$ Cep B can be tested by direct imaging.

In this paper, we present the first direct imaging detections of 
$\gamma$ Cep B (sect. 2), compare the parameters observed with the 
predictions (sect. 3), and newly determine the masses of 
$\gamma$ Cep Aa, Ab, and B (sect. 4).

\section{Imaging observations}

We observed $\gamma$ Cep first on 15 July 2006 with the Coronagraphic Imager for 
Adaptive Optics (CIAO) on the Japanese 8m telescope Subaru on Mauna Kea, Hawaii,
in the K-band. The night was photometric with $\sim 0.7$ arc sec
natural seeing. We took ten images (always 30 times 0.1 sec each,
i.e. 30 sec total integration time) 
through a semi-transparent coronographic mask (0.6 arc sec diameter,
a chrome spot on a sapphire substrate) in order to block most of the
light of $\gamma$ Cep A, but still to be able to detect B.

The data were reduced in the normal way:
Dark subtraction, flat-field devision, and adding up the
individual images. The resulting image is shown in Fig. 1.

\begin{figure}
\centering
\resizebox{\hsize}{!}{\includegraphics[clip]{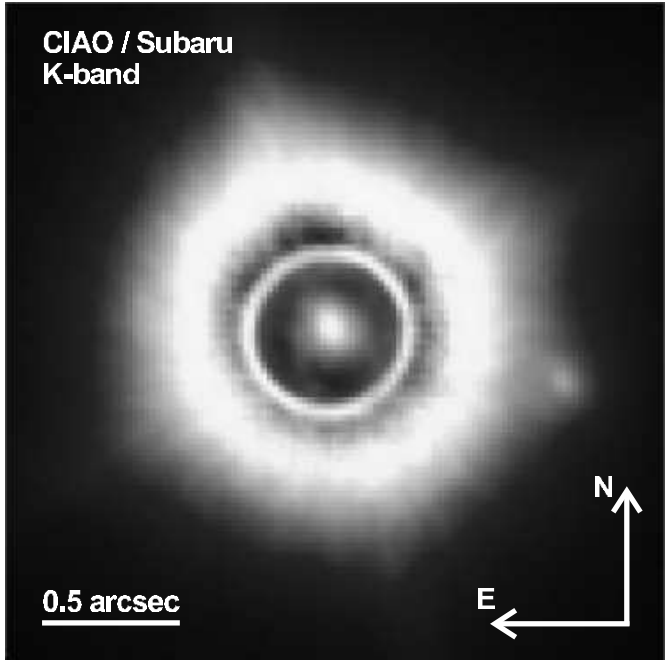}}
\caption{Subaru CIAO K-band image of $\gamma$ Cep A and B,
where A is below the semi-transparent coronograph.
The total integration time was 30 sec, obtained on 15 July 2006.}
\end{figure}

\begin{figure}
\centering
\resizebox{\hsize}{!}{\includegraphics[clip]{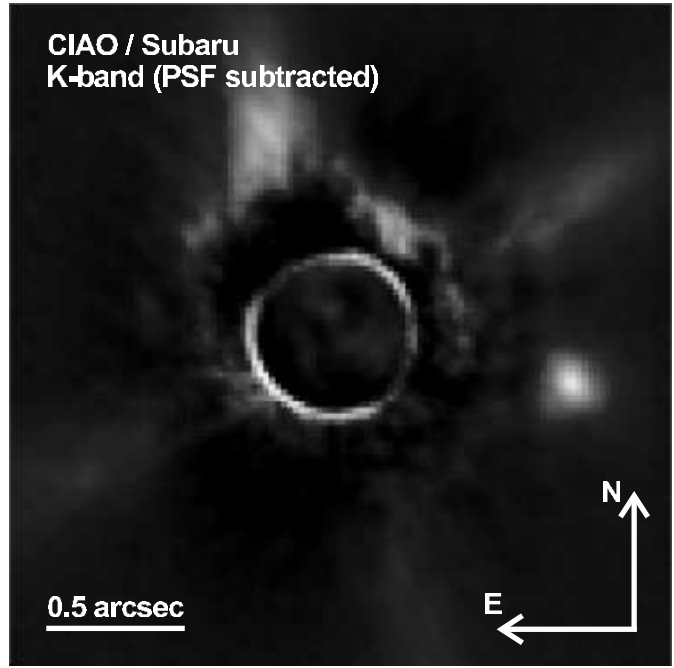}}
\caption{Subaru CIAO K-band image of $\gamma$ Cep A and B,
now after subtraction of the PSF of $\gamma$ Cep A.
The stellar companion $\gamma$ Cep B is now clearly detected 
0.87 arc sec west of A (slightly south of west).}
\end{figure}

The semi-transparent coronograph 
has a transmission of between about $1\%$ and $6\%$, 
thus still allowing the primary star to shine through.
We subtracted the 
point spread function (PSF) of remaining light of $\gamma$ Cep A, 
see Fig. 2 for the result.
$\gamma$ Cep B is clearly detected in Fig. 1 and 2.

Because the throughput of the coronograph is not well known
and because the star A may still be in the non-linear regime,
we cannot derive the magnitude of B with respect to A from
the image itself. A photometric standard star is needed.
We therefore observed the standard star FS 150 
($9.94 \pm 0.01$ mag, UKIRT web page) in the same night 
($8 \times 3$ sec) to obtain the magnitude of $\gamma$ Cep B by
aperture photometry (after PSF subtraction). 
The result for B is K = $7.32 \pm 0.11$ mag.
Hence, the magnitude difference between $\gamma$ Cep A and B
is $6.28 \pm 0.24$ mag (A has K$_{\rm s} = 1.04 \pm 0.21$ mag, 
2MASS Catalog, Skrutskie et al. 2006). 

By Gaussian centering, we can determine the position of B
relative to A, i.e. to measure the separation and position angle.
For a companion (or candidate) to a very bright primary, it is
very useful to observe with a {\em semi-transparent} coronograph,
so that the primary star is still visible together with the companion 
(for separation measurement).
For the astrometric analysis, we use the most recent pixel scale 
for CIAO, namely $21.3 \pm 0.1$ mas/pixel (Mayama et al. 2006) 
and the orientation of the detector 
being $0.81 \pm 0.25$ deg from north 
as measured by us from images of three sources in the V1686 Cyg region 
taken in the same night.

The resulting separation and position angle are listed in Table 1.

\begin{table*}
\begin{tabular}{lcccccc}
\multicolumn{7}{c}{{\bf Table 1.} Astrometry and photometry for $\gamma$ Cep A and B} \\ \hline
Obs date & Telescope  & PA & O-C & separation & O-C & $\Delta K$ \\
HJD      & instrument & $[^{\circ}]$ & $[^{\circ}]$ & [arc sec] & [mas] & [mag] \\ \hline
2453932.046 & Subaru 8.3m CIAO & $256.91 \pm 0.27$ & $-0.20$ & $0.870 \pm 0.005$ & $-0.005$ & $6.28 \pm 0.24$ \\
2453989.520 & CA 3.5m $\Omega$ Cass & $256.16 \pm 0.35$ & $+0.09$ & $0.891 \pm 0.006$ & $+0.005$ & $6.28 \pm 0.13$ \\
2453990.514 & CA 3.5m $\Omega$ Cass & $256.39 \pm 0.34$ & $+0.33$ & $0.887 \pm 0.005$ & $+0.001$ & $6.12 \pm 0.11$ \\ \hline
\end{tabular}

Remark: 
Observed minus computed (O--C) residuals are explained in sect. 3.
\end{table*}

After our detection with Subaru we observed
$\gamma$ Cep with the $\Omega$ Cass IR imager on the
Calar Alto 3.5m telescope in Spain. 
Due to the fact that our instrument of choice, the AO system ALFA, 
was not functioning at the time,
we observed $\gamma$ Cep in a backup programme by taking several thousand 
short (0.1 sec) integration images (speckle), which we then combined by shift+add,
after background correction and flatfielding.
We detected $\gamma$ Cep B on both 
11 Sept (7391 images) and on 12 Sept 2006 
(18140) using the $K^{\prime}$ (K-prime) filter
together with Br $\gamma$.
In addition to measuring the separation between B and
A from the resulting images in each night, we determined also the
magnitude difference relative to the primary, which was neither
saturated nor in the non-linear regime of the detector.
See Fig. 3 for 
the final images of the two nights
and Table 1 for the results of our measurements.
We used the Hipparcos multiples HIP 109474, HIP 91115, HIP 92961, HIP 104536,
observed in the same two nights,
for astrometric calibration of the pixel scale 
($38.681 \pm 0.059$ mas/pixel) and the detector position angle
($21.362 \pm 0.037 ^{\circ}$).

\begin{figure*}
\centering
\resizebox{\hsize}{!}{\includegraphics[clip]{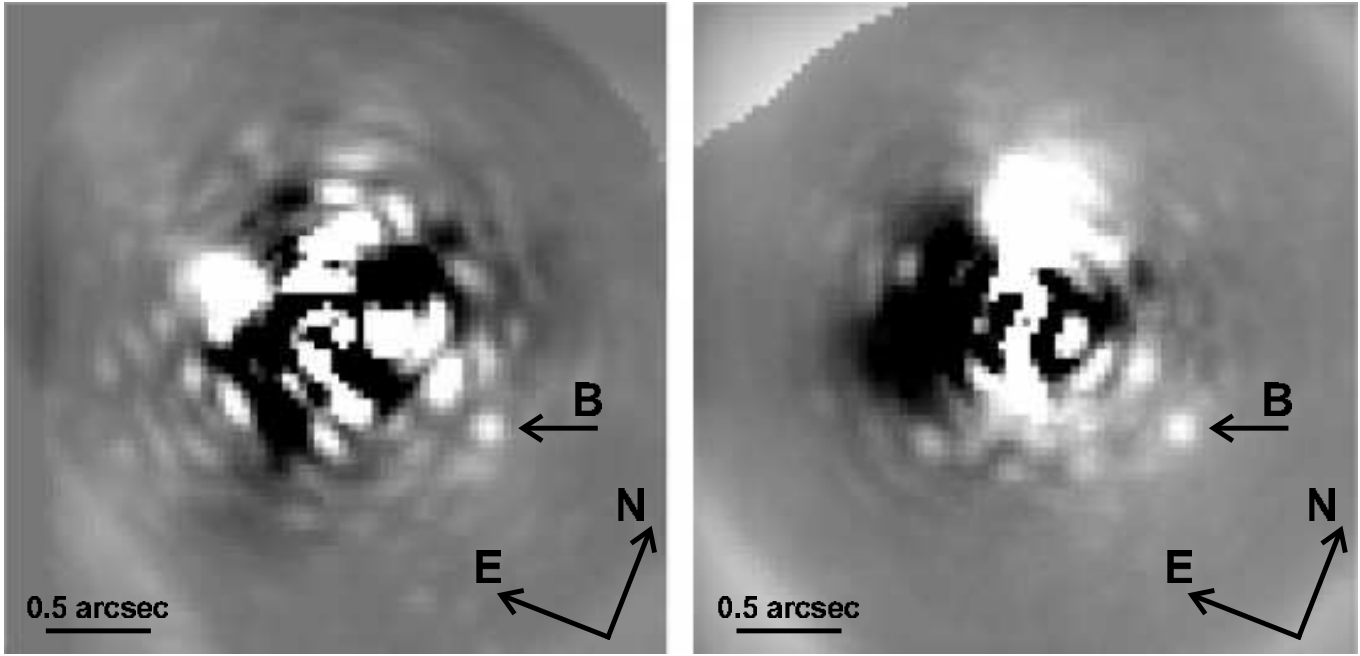}}
\caption{$\Omega$ Cass IR speckle images of $\gamma$ Cep A and B,
obtained on 11 Sept 2006 (left) and 12 Sept 2006 (right) 
with the Calar Alto 3.5m telescope with $\sim 12$ to 30 min
total integration time (through K$^{\prime}$ and Br $\gamma$ filters).}
\end{figure*}

Our measurements may indicate a small change in the position angle of
$0.63 \pm 0.36$ deg between July and September 2006, and a more
significant increase in separation of $19 \pm 6$ mas, 
consistent with the predictions for orbital motion, see Fig. 4.

For the secondary star $\gamma$ Cep B, imaged for the first time here,
we obtain a magnitude of $K=7.3 \pm 0.2$ mag,
the average of the Subaru and the two Calar Alto data points.

\begin{figure}
\centering
\resizebox{\hsize}{!}{\includegraphics[clip]{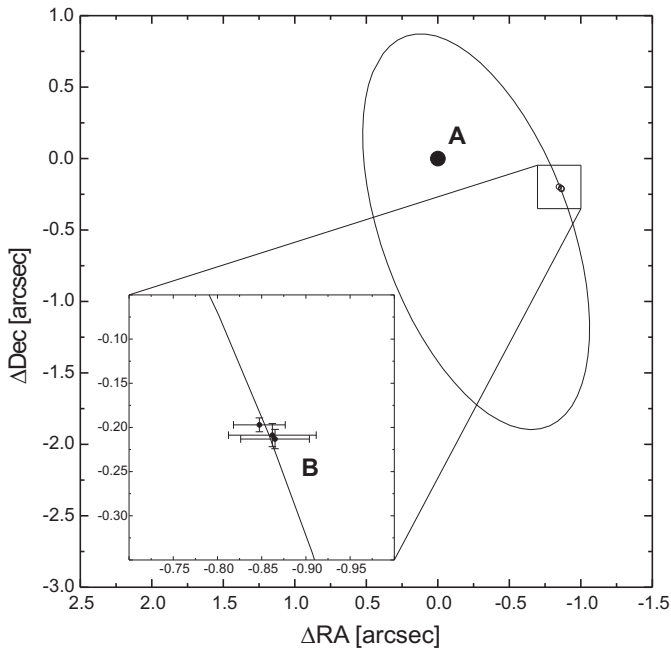}}
\caption{Orbit and orbital motion of $\gamma$ Cep B around A as
already observed by our imaging within two month.
The inlay shows our three imaging data points of B in orbit around A.
}
\end{figure}

\section{Comparison with predictions}

Our new measurements of the relative position of $\gamma$~Cep~B with
respect to the primary star allow us to test the predictions from the
orbital solution for the binary produced recently by Torres (2006). 
That study combined all available measurements of the radial
velocity of the primary from a variety of sources (including the
high-precision studies that uncovered the sub-stellar companion, along
with classical measurements from the literature and from new
observations) and used also the intermediate astrometry from the {\it
Hipparcos\/} mission (`abscissa residuals') as well as ground-based
astrometry reaching back more than a century. This enabled the scale
and orientation of the binary orbit to be established for the first
time, in a combined solution that included also the perturbing effects
of the sub-stellar companion.

A comparison of our measurements with the published orbit (Torres 2006) indicates
systematic (observed minus computed) $O-C$ residuals 
of $\sim 8^{\circ}$ in position angle and
50--60 mas in separation (positive in both cases). 
These are many times larger than the measurement errors, 
although the predicted positions themselves have 
uncertainties of about 3.1 degrees in position angle 
and 34 mas in separation for the dates of our observations. 
These reflect the uncertainty in the orbital elements as
well as the estimated error of the adopted primary mass, 
which was external to the orbital solution of Torres (2006) 
and was established from stellar evolution considerations. 
Therefore, the residuals correspond to 1.5--2.5$\sigma$, 
which, while not overly large, may nevertheless indicate the 
need for some adjustment in the orbit given
that they are systematic in nature. 
The new $O-C$ values given in Table 1 (this paper) are computed for
the new orbit (Table 2) after taking into account our three
imaging observations. They are much smaller than in Torres (2006),
i.e. a clear improvement as expected, which was the motivation
for the imaging observations.
The relative brightness of the
secondary was predicted by Torres (2006) to be $\Delta K \sim 6.4$
mag, which is not far from the values we measure (see Table 1).

\section{Improved masses}

The assumption of a value for the primary mass in the Torres (2006)
solution was necessary because of the lack of direct measurements of
the relative position between $\gamma$~Cep A and B, since the
secondary had never been imaged. With our present results this
assumption can now be dropped, and this allows for a direct
measurement of the dynamical masses of both stars with no need for
external constraints. In order to update the orbital solution and
solve for the masses we have added to the data considered by Torres
(2006) our 3 measurements of the separation and position angle of
$\gamma$~Cep with their corresponding uncertainties. We have modified
the procedures to include the relative semimajor axis of the binary,
$a''_{\rm AB}$, as an adjustable parameter in addition to the other
variables considered in the original solution. The results are shown
in Table 2, which includes the most relevant elements of the inner
and outer orbits and several derived quantities. The notation follows
that of Torres (2006). Most of the orbital elements change very little
(well within the errors). The most significant change is in the
position angle of the ascending node ($\Omega_{\rm AB}$), which
increases by about $5 ^{\circ}$. This accounts for a large fraction of
the residuals in P.A.\ noted above. The other significant change is in
the scale of the orbit, which increases by about $6\%$. This in turn
accounts for the residuals in the angular separation. The 
residuals from this new orbit are listed in Table 2, and are now all
within 1$\sigma$, indicating a satisfactory fit.

\begin{table}
\begin{tabular}{ll}
\multicolumn{2}{c}{{\bf Table 2.} Global orbital solution for $\gamma$ Cep} \\ \hline
\multicolumn{2}{c}{Parameter and value} \\ \hline
\multicolumn{2}{c}{Adjusted quantities from outer orbit (A+B)} \\
~~~~$P_{\rm AB}$ (yr)\dotfill               &  67.5~$\pm$~1.4 \\
~~~~$\gamma$ (km/s)\dotfill                 &  $-$42.943~$\pm$~0.046 \\
~~~~$K_{\rm A}$ (km/s)\dotfill              &  1.932~$\pm$~0.014 \\
~~~~$e_{\rm AB}$\dotfill                    &  0.4112~$\pm$~0.0063 \\
~~~~$\omega_{\rm A}$ (deg)\dotfill          &  161.01~$\pm$~0.40 \\
~~~~$T_{\rm AB}$ (yr)\dotfill               &  1991.605~$\pm$~0.031 \\
~~~~$a''_{\rm AB}$ (arc sec)\dotfill        &  1.467~$\pm$~0.046 \\
~~~~$a''_{\rm A}$ (mas)\dotfill             &  332.4~$\pm$~7.7 \\
~~~~$i_{\rm AB}$ (deg)\dotfill              &  119.3~$\pm$~1.0 \\
~~~~$\Omega_{\rm AB}$ (deg)\dotfill         &  18.04~$\pm$~0.98 \\ \hline
\multicolumn{2}{c}{Adjusted quantities from inner orbit (Aa+Ab)} \\
~~~~$P_{\rm A}$ (days)\dotfill              &  902.9~$\pm$~3.5 \\
~~~~$K_{\rm Aa}$ (m/s)\dotfill              &  27.0~$\pm$~1.5 \\
~~~~$e_{\rm A}$\dotfill                     &  0.115~$\pm$~0.058 \\
~~~~$\omega_{\rm Aa}$ (deg)\dotfill         &  63~$\pm$~27 \\
~~~~$T_{\rm A}$ (HJD$-$2,400,000)\dotfill   &  53147~$\pm$~71 \\ \hline
\multicolumn{2}{c}{Derived quantities} \\
~~~~$\mu_{\alpha}^*$ (mas~yr$^{-1}$)\dotfill  &  $-$63.86~$\pm$~0.98 \\
~~~~$\mu_{\delta}$ (mas~yr$^{-1}$)\dotfill    &  +150.76~$\pm$~0.43 \\
~~~~$\pi$ (mas)\dotfill                       &  72.69~$\pm$~0.41 \\
~~~~$a_{\rm AB}$ (AU)\dotfill                 &  20.18~$\pm$~0.66 \\
~~~~$M_{\rm Aa}$ (M$_{\sun}$)\dotfill         &  1.40~$\pm$~0.12 \\
~~~~$M_{\rm B}$ (M$_{\sun}$)\dotfill          &  0.409~$\pm$~0.018 \\
~~~~$f(M_p)$ ($10^{-9}$~M$_{\sun}$)\dotfill   &  1.81~$\pm$~0.31 \\
~~~~$M_p \sin i$ (M$_{\rm Jup}$)\dotfill      &  1.60~$\pm$~0.13 \\ 
~~~~$a_{{\rm Aa}-p}$ (AU)\dotfill             &  2.044~$\pm$~0.057 \\ \hline
\end{tabular}

Remark: The symbols have the same meaning as in Torres (2006).
\end{table}

The resulting masses of the two stars are $M_{\rm Aa} = 1.40 \pm
0.12$~M$_{\odot}$ and $M_{\rm B} = 0.409 \pm 0.018$~M$_{\odot}$.  Both
the primary and secondary are somewhat more massive than assumed by
Torres (2006). Consequently the minimum mass of the sub-stellar
companion also increases to $M_p \sin i = 1.60 \pm 0.13$~M$_{\rm
Jup}$, a value intermediate between those proposed by Torres (2006)
and Hatzes et al. (2003),
and the upper mass limit from Torres (2006) changes 
from 17~M$_{\rm Jup}$ to 19~M$_{\rm Jup}$.
The brightness of the stellar companion B is fully consistent with
a single M4-type dwarf, neither multiple itself nor a white dwarf.

Given the new parameters in Table 2, the critical semi-major axis for 
long-term stability of planets according to the Holman \& Wiegert (1999) 
criterion is at only $3.85 \pm 0.38$ AU. 
The size of a planet-forming circumstellar disk in the close binary is
not larger than 3.4 AU according to the Pichardo et al. (2005) criteria.
The planet candidate $\gamma$ Cep Ab orbits around Aa between 1.8 and 2.3 AU,
which is inside the stable region, so that it is not neccessary
to invoke migration after its formation.

Continued measurements of the relative position between $\gamma$~Cep A
and B over the coming years, which we intend to carry out, 
should further improve the stellar mass determinations and help 
establish also the true mass of the sub-stellar companion.

\begin{acknowledgements}
We would like to thank the staff at the Subaru 
and Calar Alto telescopes for their help during 
our observations as well as Kengo Tachihara for his support.
GT acknowledges partial support for this work from NSF grant
AST-0406183 and NASA Origins grant NNG04LG89G.
TS acknowledges support from a Thuringian State Scholarship (until July 2006) 
and from a Scholarship of the Evangelisches Studienwerk e.V. Villigst 
(since August 2006).
\end{acknowledgements}

\end{document}